\documentstyle[11pt,aaspp4]{article}
\tighten 
\begin{document}
\newcommand{\lya}{Lyman~$\alpha$}
\newcommand{\lyb}{Lyman~$\beta$}
\newcommand{\za}{$z_{\rm abs}$}
\newcommand{\ze}{$z_{\rm em}$}
\newcommand{\cmtwo}{cm$^{-2}$}
\newcommand{\nhi}{$N$(H$^0$)}
\newcommand{\degpoint}{\mbox{$^\circ\mskip-7.0mu.\,$}}
\newcommand{\kms}{\,km~s$^{-1}$}      
\newcommand{\minpoint}{\mbox{$'\mskip-4.7mu.\mskip0.8mu$}}
\newcommand{\peryr}{\mbox{$\>\rm yr^{-1}$}}
\newcommand{\secpoint}{\mbox{$''\mskip-7.6mu.\,$}}
\newcommand{\sqdeg}{\mbox{${\rm deg}^2$}}
\newcommand{\squig}{\sim\!\!}
\newcommand{\subsun}{\mbox{$_{\twelvesy\odot}$}}
\newcommand{\et}{{\rm et al.}~}

\def\ltsima{$\; \buildrel < \over \sim \;$}
\def\simlt{\lower.5ex\hbox{\ltsima}}
\def\gtsima{$\; \buildrel > \over \sim \;$}
\def\simgt{\lower.5ex\hbox{\gtsima}}
\def\arcs{$''~$}
\def\arcm{$'~$}
\def\erf{\mathop{\rm erf}}
\def\erfc{\mathop{\rm erfc}}
\title{LYMAN CONTINUUM EMISSION FROM GALAXIES AT ${\rm \bf z \simeq 3.4}$
\altaffilmark{1}}
\author{\sc Charles C. Steidel\altaffilmark{2}}
\affil{Palomar Observatory, California Institute of Technology, MS 105--24, Pasadena, CA 91125}
\author{\sc Max Pettini}
\affil{Institute of Astronomy, University of Cambridge, Madingley Road, Cambridge CB3 0HA, UK}
\author{\sc Kurt L. Adelberger}
\affil{Palomar Observatory, California Institute of Technology, MS 105-24, Pasadena, CA 91125}

\altaffiltext{1}{Based on data obtained at the 
W.M. Keck Observatory, which 
is operated as a scientific partnership among the California Institute of Technology, the
University of California, and NASA, and was made possible by the generous financial
support of the W.M. Keck Foundation.
} 
\altaffiltext{2}{Packard Fellow}
\begin{abstract}
We report the detection of significant Lyman continuum flux in the composite spectrum
of 29 Lyman break galaxies (LBGs) with redshifts $\langle z \rangle = 3.40\pm0.09$. 
After correction for opacity due to
intervening absorption using a new composite QSO spectrum evaluated at the same redshift, the 
ratio of emergent flux density at 1500 \AA\ in the rest frame to that in the Lyman continuum is
$L(1500)/L(900) = 4.6 \pm 1.0$. 
If the relative intensity of the inferred escaping Lyman
continuum radiation is typical of LBGs at $z\sim 3$ (the galaxies in this
sample are drawn from the bluest quartile of LBG spectral energy distributions due
to known selection effects), then observed LBGs
produce about 5 times more H--ionizing photons per unit co-moving volume 
than QSOs at $z \sim 3$. The associated contribution to the 
metagalactic ionizing radiation field is $J_{\nu}$(912 \AA) $\approx 1.2\pm0.3\times 10^{-21}$ ergs s$^{-1}$ cm$^{-2}$
Hz$^{-1}$ sr$^{-1}$ at $ z\sim 3$, very close to most estimates of the radiation field background
based on the ``proximity effect''. 
A preliminary analysis of the density of faint QSOs in our Lyman break galaxy survey indicates
that the standard extrapolated QSO luminosity function 
may slightly over-predict the QSO contribution to $J_{\nu}(912)$ at $z \sim 3$.  
We briefly discuss the implications of a galaxy--dominated UV background at
high redshifts. 
\end{abstract}

\section{INTRODUCTION}

The question of the origin of the UV radiation field that re-ionized the universe
at $z > 6$ and maintained the ionization of the intergalactic medium
at lower redshifts has been addressed numerous times since it was first considered
20 years ago (Sargent \et 1980).  The relative contributions of
QSOs and massive stars in galaxies to the ionizing photon budget at high redshifts
has important implications for the spectral energy distribution of the ionizing
background (Miralda-Escude \& Ostriker 1990; Steidel \& Sargent 1989) which in turn
affects the equation of state of the diffuse intergalactic medium. 
Early consideration of the evolution of the diffuse intergalactic medium suggested that
the observed QSOs would have an increasingly difficult time producing enough ionizing
photons to explain the absence of an observed Gunn-Peterson (1965) effect at redshifts
$z \simgt 3$ (e.g., Donahue \& Shull 1987, Shapiro \& Giroux 1987), suggesting the need
for additional sources of ionizing photons at high redshift. 

The contribution of star-forming galaxies to the UV background was first
explicitly considered by Bechtold \et (1987), but at the time there was little 
relevant data.  Progress in very deep galaxy surveys (e.g., Songaila, Cowie, \& Lilly 1990, Steidel \& Hamilton 1993) 
suggested that high redshift ($z \simgt 3$) star-forming galaxies might well exist,
but many uncertainties in their contribution to the ionizing background remained due to the lack
of spectroscopic redshift identifications and the uncertain opacity of galaxies to their
own Lyman continuum radiation. 

The advent of the Keck telescopes in the mid-1990s 
made possible spectroscopy of large numbers of star forming galaxies at redshifts $z \sim 3$,
where the strongest constraints on the properties of the intergalactic medium were available
from QSO absorption line surveys,  where the ability of known QSOs to maintain the ionization
of the IGM became uncertain, and where the rest-frame Lyman limit of hydrogen could be
observed from the ground. This population of UV-bright star forming
galaxies (Steidel \et 1996), generally referred to as Lyman break galaxies (LBGs), is selected photometrically by 
exploiting the continuum discontinuity 
near the rest frame Lyman limit due to the intrinsic
spectral energy distribution of hot stars, Lyman continuum opacity from H~I layers in the galaxy itself,
and, most model-independently, the opacity of the intergalactic medium due to intervening
H~I (e.g., Madau 1995). While this large continuum break (only a portion of which is
expected to be intrinsic to the galaxy) makes the discovery of LBGs extremely
straightforward and statistically complete, 
it also makes the measurement of leaking Lyman continuum radiation exceedingly difficult,
even with 10-m class telescopes.   

The problem of intergalactic HI opacity can be minimized by observing the Lyman continuum in
galaxies at smaller redshifts, but this requires
observations in the vacuum UV from space. The most sensitive direct
measurement to date, by Leitherer \et (1995), used {\it Hopkins Ultraviolet Telescope}
far-UV spectra of 4 nearby
starburst galaxies to place constraints on the fraction of Lyman continuum photons escaping
the galaxies, $f_{\rm esc} < 3$\% (but see Hurwitz, Jelinsky, \& Dixon 1997). Similarly
small upper limits for relatively nearby galaxies have been set by several groups 
using less direct methods. 
Nevertheless, it is not clear how relevant these limits are to the situation at high redshift, where
galaxies which simultaneously possess very large UV luminosities and low UV extinction are much more common
(cf. Adelberger \& Steidel 2000). 

This paper presents what we believe to be the first direct detection
of Lyman continuum flux from galaxies, made possible by a high S/N composite spectrum 
formed from a relatively small subset of $z \sim 3$ LBGs that are 
at high enough redshift to bring the rest-frame Lyman limit into a region of good spectroscopic 
sensitivity for current instrumentation. This measurement, which we currently regard
as plausible, but tentative, has significant implications for the nature of the UV radiation
field at high redshifts. In the context of evaluating the significance, we also present     
some preliminary results on the contribution of faint QSOs to the $z \sim 3$ UV
background.

\section{DATA}

The spectra selected for this study were chosen from among the 875 spectra of 
$z \sim 3$ LBGs we have obtained with the Keck telescopes and Low Resolution Imaging
Spectrometer (Oke \et 1995) between 1995 October and 1999 November.  All of the
spectra were obtained using the same instrumental configuration, a 300 lines/mm
grating blazed at 5000 \AA\ and a slit mask with 1\secpoint4 slits, providing spectral
resolution of $\sim 12.5$ \AA\ in the observed frame. The typical total exposure time
was 7200 seconds.  In its current configuration, the total throughput
of LRIS with this grating is $\sim$20\% at 4000 \AA\ and $\sim 10$\% at 3800\AA;
in order to include only spectra with adequate sensitivity at the rest-frame
Lyman limit, we have confined ourselves to the small subset of our spectra
which have $z \ge 3.300$ and ${\cal R} \le 25.0$ (81 objects).  Some of these
spectra did not reach the Lyman limit on the blue side, due
to the geometry of the slit masks, and
some were obtained under poor conditions, were contaminated by light from
neighboring galaxies, or were affected by other obvious problems.
Two dimensional sky-subtracted
spectrograms of each object were inspected for such problems, and only the cleanest
looking slits with no obvious sky subtraction problems were retained. The final
sample consists of 29 spectra with $\langle z \rangle =3.40\pm0.09$; the average
properties of the LBGs are summarized in Table 1. The important observed-frame
spectral region for evaluating the Lyman continuum is 3880-4020 \AA\ at the average
redshift.  

The composite spectrum of the LBGs was constructed by shifting each extracted, one-dimensional,
flux calibrated spectrum into the rest frame and averaging after scaling to a common median in
the rest wavelength range 1250--1500 \AA. A simple rejection scheme was used in which
equal numbers of positive and negative outliers were excluded from the average at each
dispersion point (primarily in order to exclude sky subtraction residuals near bright night sky lines).
The overall nature of the co-added spectrum was found to be insensitive to the number
of points excluded from the average at each dispersion point. The final spectrum, shown
in Figure 1, results from the rejection of 2 positive and 2 negative outliers from each
dispersion point.  The composite spectrum does not depend strongly on the adopted
scheme for combining the individual spectra.

The composite spectrum clearly shows the net effects of the Lyman $\alpha$ forest shortward
of the Lyman $\alpha$ emission line, reducing the flux by a factor of $\simeq 1.8$ relative to
the flux in the $1250-1500$ \AA\ range. Also evident is the general diminution of the flux shortward
of Lyman $\beta$ due to increasing numbers of intervening and intrinsic absorption features.
Many stellar and interstellar absorption features can be recognized in the composite spectrum;
the most prominent are labeled in Figure 1. Interestingly, averaging the Lyman $\alpha$ forest
absorption over many sight-lines brings out stellar and interstellar lines which are normally
masked by the forest in any single spectrum. 
Stellar C~III $\lambda 1176$ and He~II $\lambda 1084$ (cf. Walborn \& Bohlin 1996;
Fullerton \et 2000), and interstellar Lyman $\beta$, $\gamma$, and $\delta$ are the clearest
examples.  The Lyman $\alpha$ equivalent
width in the composite spectrum in Figure 1 is 19.7\AA\ , placing it in the
top 20-25\% of LBGs in terms of Lyman $\alpha$ line strength (cf. Steidel \et 2000). 
This large Lyman $\alpha$ equivalent width results from a selection bias in our Lyman-break sample towards
bluer objects at higher redshifts; at $z\sim 3.4$ galaxies
will satisfy our color-selection criteria only if their
intrinsic broad-band colors lie in the bluest
quartile of observed LBG UV colors, and a strong contribution from Lyman-$\alpha$ emission
helps galaxies achieve the required blue broad-band color.  See Steidel \et (1999)
for a more complete discussion.
The impact of this selection effect on our present results is
considered in \S 4 below.

The bottom panel of Figure 1 shows residual flux in the Lyman continuum region.
We have chosen to evaluate the residual flux over the rest wavelength range
880-910 \AA, for two main reasons: first, the incidence of intervening Lyman limit
absorption systems (Sargent, Steidel, \& Boksenberg 1989; Stengler-Larrea \et 1995)
at $z \sim 3.4$ means that broad-band evaluation of the decrement across the Lyman
limit is not likely to provide a good estimate of the emergent spectrum from the
galaxies---the mean free path of a Lyman continuum photon at $z \sim 3$ is
only $\Delta z \simeq 0.17$ (Haardt \& Madau 1996), corresponding to $\sim 35$ \AA\ in
the rest frame at $z \simeq 3.4$. 
Second, the 880-910 \AA\ range includes {\it observed} spectral regions
that are still within the range for which LRIS provides adequate throughput
for the entire range of redshifts included in the composite spectrum. While the significance per
resolution element
of the residual flux in the Lyman continuum is low, the net flux observed over the rest-frame
30 \AA\ bandpass is formally significant at the $4.8\sigma$ level (see Table 2). 

An obvious concern in gauging the reality of the residual flux, which corresponds to an
average observed flux density of $m_{\rm AB}=27.4$ in the narrow wavelength range, is the
extent to which systematic problems with sky subtraction might mimic residual Lyman
continuum flux. We experimented extensively with the data, using a number
of different tests. First, we confirmed that the signal is not contributed by
a few outlier spectra by combining the 1-D spectra using different subsets with
different outlier rejection algorithms. We checked for overall systematics
in the 2-D sky subtraction by combining the 2-D background-subtracted slitlets
after rebinning in the wavelength direction to a common rest wavelength interval, registering
on the spatial position of the LBG, and then experimenting with various methods of
residual background subtraction (i.e., second-pass subtraction of the background that
might eliminate systematics that were too subtle to notice in any single slitlet).  
We then evaluated  
the residual flux in the appropriate wavelength range both on and off the spatial positions of
the detected continuum longward of the Lyman limit; the results are consistent with the
residual flux measured from the stacked 1-D spectra, although all of the additional
image processing necessary to combine the spectra in 2-D could itself be a source
of systematics errors. 
Probably the most direct test of sky subtraction systematics is to examine spectra,
taken with the same instrumental configuration (all observed on slit masks as part
of the general LBG survey), of similar objects 
that are known to have optically
thick Lyman limits at the same observed wavelengths as the region of interest
for the LBGs. In our spectroscopic LBG survey, we have observed 3 faint ${\cal
R}=22.0$,22.2, 23.6) QSOs
which have optically thick Lyman limit systems (confirmed subsequently with
long-slit observations at higher dispersion) in the redshift range $3.29 \le z_{LLS}
\le 3.43$. These spectra were shifted into the rest frame of the intervening
Lyman limit system, scaled, and averaged to produce the composite spectrum shown in Figure 2.
Figure 2 illustrates that
optically thick Lyman limit systems result in a measured flux in the [880,910] \AA\
interval consistent with
zero, making less likely the possibility that scattered light in the instrument
is responsible for the residual flux shortward of 912 \AA\ observed in the composite
LBG spectrum.  

In summary, none of the tests that we have performed 
gives us reason to believe that the residual flux shortward of 912 \AA\ in Figure 1
is an artifact of systematics.
Nevertheless, as discussed below, it would be prudent to view the detection with 
some skepticism until confirmed by higher quality spectra of individual objects. This is very
difficult with existing instrumentation, but we expect that LRIS-B, the blue channel for LRIS
optimized for the 3100-5500 \AA\ wavelength range and soon to be commissioned on the Keck I
telescope, will be ideally suited to making improved measurements on
the much larger numbers of bright LBGs in our sample at $z \sim 3.0$.   

\section{THE EMERGENT FAR-UV SPECTRUM OF LBGs}

One of the complications to the interpretation of the flux measurement above is that 
the opacity in the galaxies' rest-frame Lyman continuum (in fact, for all
spectral regions shortward of Lyman $\alpha$ emission) has a significant contribution from  
intervening material not directly associated with the galaxies.  
This contribution must be estimated accurately in order to
infer the emergent spectrum of the star-forming galaxies. 
A common way to accomplish this is with models of intergalactic opacity based on  
column density distributions of Lyman series lines and the incidence of higher
column density systems that contribute significant Lyman continuum opacity (e.g., Lyman
limit systems). The distribution functions, obtained from QSO absorption line surveys,
are used to estimate the opacity of the IGM as a function
of observed wavelength for objects at a given redshift (e.g., Madau 1995; Bershady, Charlton,
\& Geoffroy 1999). 
For the present purposes we have opted to use
a more direct empirical method that bypasses the need to model the opacity via fits to
observed distribution functions. We have formed
a composite QSO spectrum by combining spectra obtained for surveys of intervening 
Lyman limit systems (see Sargent, Steidel, \& Boksenberg 1989, Stengler-Larrea \et 1995)
of QSOs in the same range of redshift as the galaxies considered above. The spectra
were chosen from among more than 100 QSOs in the redshift range $2.75 \le z \le 4.10$ all observed
using the Palomar 200-inch telescope and Double Spectrograph; the spectral resolution in the
$3150-4700$ \AA\ range was 4\AA.  The composite spectrum,
which is effectively an average over 15 lines of sight for QSOs with a mean redshift
$\langle z_{\rm QSO} \rangle=3.47\pm0.14$, has been scaled to the composite LBG spectrum, rebinned
to a similar resolution, and over-plotted on Figure 1. Note the similarity of the QSO and galaxy SED over the rest
wavelength range 1050-1700 \AA, and that the Lyman $\alpha$ forest decrements in the
rest wavelength range 1050-1170 are equivalent for the QSO and galaxy spectra
(the measured value of the $D_A$ parameter [Oke \& Korycansky 1982] is $\approx 0.46$ for both). 

The observed flux densities in the arbitrary units of Figure 1 are summarized in
Table 2 for both the QSO and LBG spectra.  
To estimate the change in net opacity due to intergalactic material between
1100 \AA\ (where it is obvious in both the galaxy and QSO spectra and is dominated
by Lyman $\alpha$ line opacity from the forest) and rest-frame 900 \AA, we adopt the
intrinsic QSO spectral energy distribution shortward of 1050 \AA\ suggested by Madau, 
Haardt, \& Rees (1999), $f_{\nu} \propto \nu^{-1.8}$. The observed change in the
QSO continuum level between the [1060,1170] \AA\ interval and the [880,910] \AA\ interval
is a factor of $3.7\pm0.2$; the intrinsic component of this flux decrement is estimated from
the power law assumption to be a factor of 1.5, so that the inferred flux decrement at 900 \AA\ 
relative to that at 1100 \AA\  due to intervening material is an additional factor of $2.5$.   
Using this factor to correct the observed ratio $f[1100]/f[900]=9.4\pm2.0$ for
the galaxy spectrum, and accounting for the Lyman $\alpha$ line blanketing
in the [1060,1170] \AA\ interval and a small difference in unabsorbed flux between
1500 and 1100 \AA\ intrinsic to the galaxy SED (based on the mean color of the composite
spectrum after correction of the observed $G$ band for Lyman $\alpha$ line blanketing), 
one obtains an intrinsic flux density ratio
for the LBGs of $f[1500]/f[900] = 4.6\pm 1.0$. It is important to note that the {\it observed}
ratio is $f[1500]/f[900] \simeq 17.7$, or a contrast of 3.1 magnitudes on the AB scale.

Because of the effects of intergalactic absorption on
the observations, a positive measure of Lyman continuum flux becomes very rapidly more difficult
with increasing redshift; the observation is probably
impossible using present observational facilities by redshift $z \sim 4.5$, despite
higher system throughputs for ground-based spectroscopy---the intervening opacity is
approximately 3 times higher, the sky background is about 3 times brighter, and typical
galaxies are two times fainter, than at $z \sim 3$.    

\section{IMPLICATIONS}

The flux density ratio $f[1500]/f[900]=4.6 \pm 1.0$ obtained above is unexpectedly
large for star forming galaxies; it is similar to typical models of the UV SEDs of
star forming galaxies (e.g., Bruzual \& Charlot 1996) without {\it any} Lyman continuum
self--absorption from H~I in the galaxy. However, it must be realized that at present
there are no direct measurements of the intrinsic amplitude of the 
stellar Lyman break and we have to rely exclusively on models. The model predictions
depend sensitively on the initial mass function and stellar lifetimes at the very high mass
end, and on the age of the star formation episode. 
The amplitude of the stellar Lyman limit for most spectral synthesis models (e.g., Bruzual
\& Charlot 1996; Leitherer \et 1999) ranges from a factor of $\sim 3$ to $\sim 5.5$ over
a plausible range of ages and initial mass functions. More recent models (e.g., Schaerer 1998), 
which include
non-LTE effects, generally produce somewhat smaller break amplitudes than the LTE models
that are used by the popular spectral synthesis programs, as small as a factor of
$2.5$ for constant star formation over periods $> 10^7$ years (Smith \& Norris, private
communication) .  Adopting for the moment
the assumption that the stellar energy distribution has an intrinsic Lyman discontinuity
of a factor of 3, our result for $f[1500]/f[900]$ implies that the fraction of
UV ionizing photons escaping the galaxy is $f_{\rm esc} \simgt 0.5$. 

Our definition of $f_{\rm esc}$ differs from a definition sometimes
encountered elsewhere in the literature (e.g. Leitherer \et 1995).
By $f_{\rm esc}$ we mean the fraction of emitted 900\AA\ photons
that escapes the galaxy without being absorbed by interstellar
material divided by the fraction of 1500\AA\ photons that escapes.
This is the quantity required to calculate the ionizing radiation
emitted by Lyman-break galaxies from their well constrained
1500\AA luminosity density.  An alternate definition of
$f_{\rm esc}$ is simply the fraction of emitted 900\AA\ photons
that escapes (e.g. Leitherer \et 1995); in this definition
the 1500\AA\ normalization is omitted.  Because
perhaps only about 15--20\% of 1500\AA\ photons
escape from typical LBGs without being absorbed by dust
(e.g. Pettini \et 1998, Adelberger \& Steidel 2000),
values of $f_{\rm esc}$ calculated under these two
definitions will differ significantly.

The fraction of escaping Lyman continuum photons is likely to be highly variable 
from galaxy to galaxy, depending (at least) on the morphological details of the galaxy-scale gaseous
outflows that appear to be a ubiquitous signature of both high redshift and low-redshift
starburst galaxies (see Heckman 2000). 
As an example, consider the best 
far-UV spectrum of a galaxy (at any redshift) obtained to date, the gravitationally-lensed
LBG at $z=2.72$, MS1512-cB58 (Pettini \et 2000). 
In this case it would be surprising to find significant leakage of Lyman continuum photons, given
the extremely large covering fraction of optically thick (in the Lyman-continuum) gas
implied by the
the observed damped interstellar Lyman $\alpha$ line and several lines 
of low-ionization metallic species that are black at line center.
On the other hand, the interstellar lines in the spectrum of the composite $z=3.4$
LBG presented above are less than half the strength of those in cB58; it is possible
that this may be due to a smaller outflow covering fraction along our line of sight
(or a smaller average velocity
spread---it is  difficult to tell the difference in spectra of 12.5\AA\ resolution), in which
case a higher fraction of leaking Lyman continuum photons would be expected. 
As mentioned in \S 2, the galaxies chosen for the composite spectrum are drawn from the bluest quartile
of the observed LBG population. While on average they have observed UV luminosities similar to
the de-magnified luminosity of cB58, they are {\it a priori} more likely to be younger and/or
less dusty.  Thus, it may be that, because of the selection effect for the
high redshift tail of the $z \sim 3$ LBG distribution to be drawn from the bluest objects,
they may be those ``most likely to succeed'' in getting Lyman continuum photons
out into the intergalactic medium. 

Since there is now a published luminosity function for $z \sim 3$ LBGs to apparent
magnitudes of $R \sim 27$, 2.5 magnitudes fainter than $L^{\ast}$ (Steidel \et 1999), 
it is straightforward to calculate 
the contribution of LBGs to the general metagalactic radiation field at $z \sim 3$ under
the assumption that the composite spectrum is representative of the population as a whole.
We use the measured ratio $L(1500)/L(900) =4.6\pm1.0$  to convert the published LBG
luminosity function (evaluated at a rest frame wavelength close to 1500 \AA) into
a distribution of Lyman continuum luminosities. This calculation is identical to
that already carried out by Madau (2000), substituting a slightly different constant
of proportionality. Integrating the Schechter (1976) function over all observed luminosities (down
to 0.1$L^{\ast}$)
we obtain a co-moving emissivity at 1 Ryd of $1.2\pm0.3 \times 10^{26}h$ ergs s$^{-1}$ Hz$^{-1}$
Mpc$^{-3}$ (for an Einstein de-Sitter cosmology) at $z \sim 3$, 
exceeding (according to Madau, Haardt, \& Rees 1999) the 
contribution from QSOs by a factor of about 5.  Ignoring any
contribution to the radiation field from ``re-processed'' recombination and
2-photon radiation (cf. Haardt \& Madau
1996), and assuming
an effective absorption distance of $\Delta z \approx 0.17$ for a Lyman continuum photon
at $z \sim 3$, 
the implied contribution of the LBGs to the
metagalactic radiation field intensity is $J_{\nu}(912) \sim 1.2\pm 0.3 \times 10^{-21}$
ergs s$^{-1}$ cm$^{-2}$ Hz$^{-1}$ sr$^{-1}$. 
This is very close to the value typically obtained from the
QSO ``proximity effect'' (Bajtlik, Duncan, \& Ostriker 1989, Scott \et 2000, and references
therein). 
Estimates of $J_{\nu}(912)$ from the proximity effect are subject to a variety of systematic 
uncertainties (see discussion in Rauch \et 1997); independent upper
limits have been placed on the value of $J_{\nu}(912)$ based on the
lack of detection of fluorescent Lyman $\alpha$ emission from Lyman limit absorption 
clouds (Bunker \et 1999), $J_{\nu}(912) < 2 \times 10^{-21}$, but the validity
of this limit is somewhat model-dependent. Given the inherent
uncertainties in both our crude calculation and in the measurements of $J_{\nu}(912)$,
we do not believe that there is currently reason to be concerned about over-producing the UV background.

There has been extensive speculation in the literature as to whether QSOs
are sufficient to provide the implied value of $J_{\nu}(912)$ at high redshift (e.g.,
Donahue \& Shull 1987, Shapiro \& Giroux 1987,
Haardt \& Madau 1996, Scott \et 2000, Meiksin \& Madau 1993, Madau, Haardt, \& Rees 1999).
Most of the recent studies conclude that QSOs are marginally able to produce the proximity
effect background at $z \sim 3$, but begin to have a serious problem by $z \sim 4$ due
to the observed decreasing space density of bright QSOs (assumed to apply to faint QSOs as
well) that has been observed by several groups (Kennefick \et 1995, Schmidt \et 1995, SDSS
collaboration).  At $z \sim 3$, we have perhaps the best data for testing whether the 
QSO luminosity function at apparent magnitudes fainter the $R \sim 21$ is consistent with
the extrapolated luminosity functions used in computing the QSO contribution to the background
\footnote{Unlike the QSO contribution, the LBG contribution to the metagalactic UV radiation
field is based on a luminosity function where the full range of luminosities has been observed.}. 
For the LBG survey at $z \sim 3$ ($\sim 0.3$ square degrees) our observational selection of 
Lyman break objects should be at least as complete for QSOs as
it is for galaxies, since no attempt has been made to remove stellar objects from the sample, 
the SEDs of the galaxies and QSOs are quite similar, and the spectroscopic identification of
objects with broad emission lines is more straightforward than for typical LBG spectra.
Adopting the QSO luminosity function of Pei (1995), 
we should have found a total of $\approx 12$ spectroscopically identified 
QSOs in the redshift range $2.5 \simlt z \simlt 3.5$ over
the apparent magnitude range $20.5 \le {\cal R} \le 25.5$, taking into account the spectroscopic
completeness and effective
survey volumes as a function of apparent magnitude and redshift as described by Steidel \et (1999). 
Our sample of 875 spectroscopically identified objects contains a total of 8 QSOs with
$2.5 \le z \le 3.5$, 
ranging in apparent magnitude from ${\cal R}=20.6-24.8$. 
While a more careful analysis of the observed QSO density is beyond the scope of this letter, the point is
that the QSO contribution to the metagalactic background is certainly no higher than
has been estimated in recent analyses (cf. Haardt \& Madau 1996). 
Steeper faint end luminosity function slopes for QSOs at $z \sim 3$ are probably excluded by our survey
data (cf. Haiman, Madau, \& Loeb 1999). 

\section{SUMMARY AND DISCUSSION}

We have presented observational evidence for significant Lyman continuum emission
over the rest-frame wavelength range 880-910 \AA\ 
in the composite spectrum of Lyman break galaxies at $z\simeq 3.4$. After accounting
for the effects of absorption due to intervening material in the spectral
range of interest, we find an emergent flux density ratio $f(1500)/f(900) = 4.6\pm 1.0$.
This ratio has been used to convert the observed far-UV luminosity function of LBGs
into the luminosity density of Lyman continuum photons at $z \approx 3$. Under
the assumption that this spectral energy distribution is characteristic of the whole
LBG population, then LBGs provide a factor of $\sim$ 5 times more hydrogen ionizing
photons than the estimated contribution from QSOs at the same redshifts. We have 
emphasized that the LBGs comprising the composite spectrum are drawn from the
bluest quartile of intrinsic far-UV colors, so that they may be more likely than
typical LBGs to exhibit significant escaping Lyman continuum emission. However, adopting the
steep $\alpha=-1.6$ far-UV luminosity function observed for LBGs (Steidel \et 1999), intrinsically
faint objects contribute a substantial fraction of the UV luminosity density. It may
be that these faint objects are bluer on average than the $\sim L^{\ast}$ 
objects which comprise most of the LBG spectroscopic samples (cf. Meurer, Heckman, \& Calzetti 1999),
in which case the composite spectrum may be a good representation of the objects which
dominate the UV luminosity density. 
Even if the distribution of rest--frame colors is independent of UV luminosity (in which
case the composite spectrum would be representative of only $\sim 25$\% of the LBGs), 
{\it the implication is that LBGs contribute at least as many ionizing photons as QSOs at $z \sim 3$.}
We have presented preliminary results on the QSO density at faint magnitudes which emphasize
that, if anything, the QSO contribution has been slightly over-estimated. 

A metagalactic radiation field with a contribution from young galaxies that is at least
as large (at $z \sim 3$) as that from QSOs makes it much easier to understand how the intergalactic
medium can remain highly ionized at redshifts $z \simgt 4$, where the QSO contribution
to the radiation field decreases significantly (cf. Madau, Haardt, \& Rees 1999). 
While the UV luminosity function of LBGs is relatively poorly constrained at $z \sim 4$, and
even more poorly constrained at $z \simgt 5$, there are indications that the co-moving density
of star forming galaxies 
changes much more slowly than that of the QSOs at these redshifts (cf. Steidel \et 1999),
making it quite plausible that the universe is re-ionized by stars rather than AGN, and
that stars play at least as important a role as QSOs in maintaining the high ionization level
at later epochs. 

While there is much to discuss about the possible implications of a metagalactic
ionizing radiation field dominated by massive stars at $z \ge 3$, we suggest that
the results above be treated as preliminary until high quality observations of
individual galaxies exhibiting clear evidence for Lyman continuum photon leakage
become available. In particular, it will be very interesting to explore the 
variation of Lyman continuum leakage as a function of other
galaxy properties (e.g, UV color, strength of stellar and interstellar absorption lines,
UV or bolometric luminosity, etc.) as these are all factors of relevance for understanding
the physics of rapidly star-forming galaxies at high redshifts, and their importance
in affecting the hydrodynamics and chemical enrichment of the intergalactic medium.
The limitation of the current data is largely instrumental:  most spectrographs
are blind in the near-UV.  A new generation of
near-UV optimized spectrographs on 8-10m telescopes should allow wholesale
observations of the Lyman continuum region among the large existing samples of 
UV-bright galaxies at $z \sim 3$.  

\bigskip
\bigskip
We would like to thank our collaborators in the LBG survey project, Mark Dickinson,
Mauro Giavalisco, and Alice Shapley, who played important roles in making this work possible. 
Informative discussions with Raul Jimenez, Linda Smith, Danny Lennon, and Stephane Charlot 
are gratefully acknowledged. CCS and KLA have been supported by grant
AST95-96229 from the U.S. National Science Foundation and by the David and Lucile
Packard Foundation.

\bigskip
\newpage
\begin{deluxetable}{lcc}
\tablewidth{0pc}
\scriptsize
\tablecaption{Properties of LBGs Comprising the Composite Spectrum}
\tablehead{
\colhead{} & \colhead{Mean Values}  & \colhead{Range of Values} 
} 
\startdata
Redshift & $3.40\pm0.09$ & [3.300:3.648] \nl
${\cal R}$\tablenotemark{a} & $24.33 \pm 0.38$ & [23.50:24.86] \nl
$G-{\cal R}$\tablenotemark{a} & $0.89 \pm 0.18$ & [0.52:1.16] \nl
$U_n-G$\tablenotemark{a,b}   & $2.73\pm0.44$ & [1.94:3.59] \nl 
E(B--V)\tablenotemark{c} & 0.025 (mean) & [$-$0.07:0.12] \nl 
 & 0.038 (median) &  \nl
${\rm W_0(Ly \alpha)}$ & 19.7 \AA\  & [$-$20:47] \AA\ \nl
\enddata
\tablenotetext{a}{AB magnitude system; effective wavelengths are 3650, 4730, and 6830 \AA\, for
$U_n$, $G$, and ${\cal R}$, respectively
(see also Steidel \& Hamilton 1993)} 
\tablenotetext{b}{1 $\sigma$ lower limits from local background fluctuations in an isophotal aperture 
} 
\tablenotetext{c}{Inferred reddening E(B-V) relative to template spectrum, assuming
Calzetti (1997) attenuation curve  (see Steidel \et 1999, Adelberger \& Steidel 2000).
The median value for the full sample of $z \sim 3$ LBGs is $E(B-V)_{\rm med} \simeq 0.12$ }
\end{deluxetable}
\begin{deluxetable}{lcc}
\tablewidth{0pc}
\scriptsize
\tablecaption{Measurements of Composite FUV LBG and QSO Spectra\tablenotemark{a}}
\tablehead{
\colhead{} & \colhead{LBGs}  & \colhead{QSOs} 
} 
\startdata
f[880,910] & $0.19\pm0.04$ & $0.50\pm0.03$ \nl
f[1060,1170] & $1.79\pm0.03$ & $1.86\pm0.02$ \nl
f(1100)/f(900) [uncorrected] & $9.4\pm2.0$ & $3.7\pm0.2$ \nl
f(1100)/f(900) [corrected] & $3.8\pm0.8$ & $1.5\pm0.1$ \nl
f(1500)/f(900) [uncorrected] & $17.7\pm3.8$ & $6.8\pm0.4$ \nl
f(1500)/f(900) [corrected]  & $4.6\pm1.0$ & $1.9\pm0.1$ \nl
\enddata
\tablenotetext{a}{All fluxes are in the arbitrary flux density units of Figure 1. Corrected
quantities are those for which the statistical effects of intervening absorption have
been removed using the procedure outlined in the text.}
\end{deluxetable}
\newpage
\begin{figure}
\figurenum{1}
\plotone{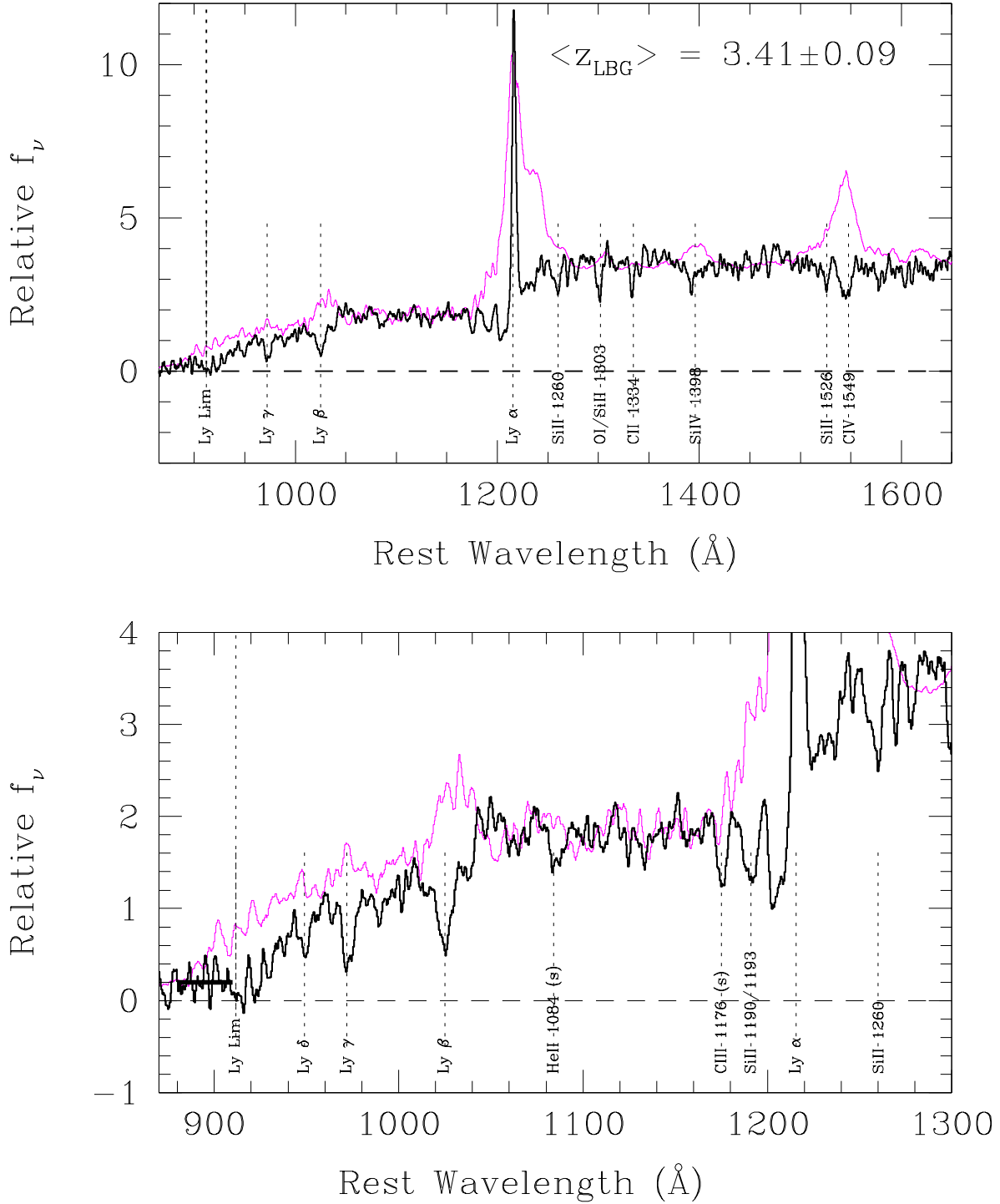}
\caption{Composite spectrum of Lyman break galaxies at $\langle z \rangle =3.40$ (dark
histogram), together
with a composite QSO spectrum drawn from QSOs over the same range of redshifts (light
histogram), as described in
the text. The spectrum has been boxcar smoothed by 1 resolution element. The position of the rest frame Lyman limit is indicated with a vertical dotted line.
Note the average residual flux in the Lyman continuum, evaluated from 880-910 \AA\ in the rest frame,
indicated with the dark horizontal line segment in the bottom panel. Positions of some notable
interstellar and stellar absorption features are indicated. Stellar features with no possible
interstellar contributions are indicated with (s). }
\end{figure}
\begin{figure}
\figurenum{2}
\plotone{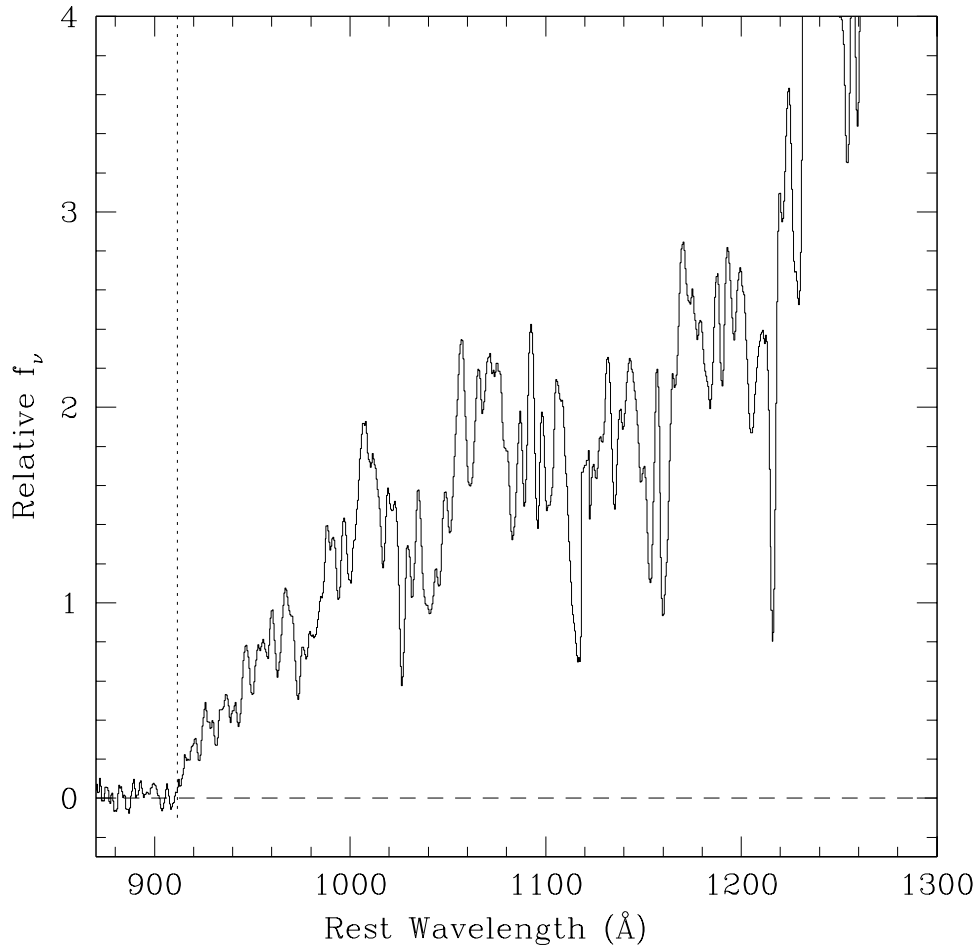}
\caption{Composite spectrum of 3 QSOs with optically thick Lyman
limit systems at $z \simeq 3.30$, combined after shifting into the rest frame of
the absorption system. The faint (${\cal R}=22.1-23.6$) 
QSO spectra were obtained using the same observational
setup as used for the galaxies comprising the composite LBG spectrum shown in 
figure 1. The residual flux in the Lyman continuum
is $0.013\pm0.018$ (in the arbitrary flux units shown in the figure) 
evaluated over the bandpass [880,910] \AA\ in the rest frame; this value is consistent
with zero, as expected.}  
\end{figure}
\end{document}